\documentclass[aps,twocolumn,floatfix]{revtex4}

\usepackage{color}
\usepackage{amssymb}
\usepackage{amsmath}
\usepackage{epsfig}
\usepackage{graphics}
\usepackage{graphicx}
\usepackage{dcolumn}
\usepackage{subfig}
\usepackage{bm}
\usepackage{float}

\def\be{\begin{equation}}
\def\ee{\end{equation}}
\def\bea{\begin{eqnarray}}
\def\eea{\end{eqnarray}}
\def\bfg{\begin{figure}[H]}
\def\efg{\end{figure}}

\begin{document}

\title{Exotic signature of dynamical quantum phase transition in the time evolution of engineered initial state}

\author{Sirshendu Bhattacharyya$^{1}$} 
\email{sirs.bh@gmail.com}
 \author{Subinay Dasgupta$^2$}

\address{$^1$Department of Physics, R.R.R Mahavidyalaya,
Radhanagar, Hooghly 712406, India}

\address{$^2$Department of Physics, University of Calcutta,
92 Acharya Prafulla Chandra Road, Kolkata 700009,
India}


\begin{abstract}
\noindent
Dynamical phase transition in quantum many body systems is usually studied by taking it in the ground state and then quenching a parameter to a new value. We investigate here the dynamics when one performs the time evolution of a generic state and observe that the rate function related to the Loschmidt echo shows non-analytic behavior of two types, one related, and the other unrelated, to the appearance of a quantum phase transition. Specifically, we consider a quantum Ising chain in an initial configuration which is a generic superposition of the eigenstates, and follow its dynamics under the transverse Ising Hamiltonian with constant field. Depending on the  the configuration of the initial state, some singularities appear in the rate function which do not correspond to the equilibrium phase transition of the Hamiltonian. However another class of singularity is found having connection with the quantum critical point. Some features of the singularities of the rate function have been derived analytically and it is observed that the occupancy of quasiparticle eigenstates plays the key role in triggering the non-analyticities.
\end{abstract}

\maketitle

\section{Introduction}
\label{intro}

\noindent

Quantum many body system at low temperatures has some unique features that gives rise to interesting physics. Experimental advancements with the ultracold atoms or traped ions have reinforced the investigation in this segment \cite{Fried-Expt,Buluta-Expt,Kim-Expt,Lew-Expt}. One of the attributes that becomes conspicuous in the low temperature limit is the quantum fluctuation, which is again the root of the phenomena of quantum phase transition (QPT) \cite{Vojta,Sachdev}. This kind of phase transition triggers an abrupt change in the ground state of a quantum many body system. The appearance of a QPT is associated with non-analytic behavior of some observables as well in the ground state. An intriguing issue is whether this behavior leaves a footprint in the dynamics of the system when it is out of equilibrium \cite{SB2,SB3}. To this end, extensive studies have been made to analyze the effect of a quench on the dynamics of some response function \cite{Pol-Rev,Ryl}. For example, suppose a Hamiltonian ${\mathcal H}(p)$ with some control parameter $p$, shows a quantum phase transition at $p=p_c$. We take the system in the ground state at $p=p_1$, quench it to $p=p_2$ and observe thereafter some measurable quantity as a function of time. Will the temporal behavior of that quantity depend crucially on whether the interval $[p_1, p_2]$ includes or excludes $p_c$? It has been reported that, for a quantum many body system the answer is `yes' in many cases \cite{Jacek-Rev}. One crucial issue is of course the choice of a suitable measurable quantity. 

In some systems it has been observed that, following a quench, the return probability of the time-evolved wave function to its initial state bears a signature if a quantum critical point (QCP) has been crossed and it can be captured by the non-analyticity of a quantity called rate function. It was first demonstrated by Heyl \emph{et al.} \cite{Heyl-PRL} in a transverse-field Ising chain quenched across the QCP. The singularities were called dynamical quantum phase transition (DQPT) points. This seminal work influenced several investigations on many other systems in search of DQPT \cite{Heyl2,Heyl2015,Andraschko,Vosk,Vajna2015,Lang,Lang-PRL,Jafari-PRB,UB,Zhou,Zvyagin,Heyl-Review1,Heyl-Review2}. For transverse Ising system, the DQPT was also observed in experiments with trapped ions and optical lattices \cite{Jurcevic,Flas-Expt}. For this system it has been observed that the order parameter, namely the longitudinal magnetization shows an oscillatory decay for quench across the QCP. The periodicity of this oscillation matches with that of the DQPT points also \cite{Heyl-PRL,Heyl-Review1,Zvyagin}. This observation suggests that DQPT has a close connection with equilibrium quantum phase transition. However, it has been demonstrated later that, in some cases the singularities show up even when the quench does not pass through a critical point and conversely, quench across a critical point does not associate with it any singularity in rate function \cite{Vajna,Jafari,Sharma,Haldar}.

Phase transitions in dynamical systems generally fall in two categories \cite{Zunkovic,Lang,Lang-PRL,Zvyagin}: one is a Landau-type transition where the long-time average of the order parameter serves as the dynamical order parameter \cite{Mori} and the other, which we have already introduced as DQPT, has no so called order parameter. The latter is generally observed following a quench in a parameter in the Hamiltonian. However, our present study finds that DQPT can occur in the unitary time evolution of an engineered initial state also.

The Loschmidt amplitude of a quantum system evolving from a state $|\Psi (0)\rangle$ is defined as $ \mathcal{G}(t) = \langle \Psi (0)|\Psi (t)\rangle$, where $|\Psi (t)\rangle$ represents the wave function of the system at time $t$. The probability of the system returning to its initial state, $ \mathcal{L}(t) = |\langle \Psi (0)|\Psi (t)\rangle|^2$ is known as the Loschmidt echo. Another related quantity $ \mathcal{R}(t) = - \lim\limits_{N\to\infty} \ln |\mathcal{G}(t)|^2/N $ is called the rate function which indicates DQPT by its non-analytic signature along the time axis. The idea of DQPT has also an analogy with that of the temperature driven phase transition of a quantum system having canonical partition function $\mathcal{Z} (\beta) = \text{Tr}\: [\exp (-\beta \mathcal{H})]$, where $\mathcal{H}$ is the Hamiltonian of the system and $\beta$ is the inverse temperature.  Writing return probability as $ \mathcal{L}(t) = |\langle \Psi (0)| \exp(-i\mathcal{H} t /\hbar |\Psi (0)\rangle|^2$, it is obvious that zeros of $\mathcal{L}(t)$ (which should be connected to singularity of the rate function) occur when the real part of the complex function $\mathcal{Z}(z) = \langle \Psi(0) | \exp (-z\mathcal{H}) | \Psi(0) \rangle $, called the dynamical partition function, vanishes and $\mathcal{Z}$ lies on the imaginary (time) axis of the $z$-plane. The zeros of Re[$\mathcal{Z}(z)$] are called Fisher zeros.
 
As mentioned earlier, whether DQPT is fundamentally connected with the equilibrium quantum phase transition or not is now a debated issue. In search of this answer, another aspect which is the role of the initial state in this phenomena becomes important. Only a few works to our knowledge have dealt with excited initial states in this regard \cite{Andraschko,Wang,Heyl2017} including the study of Loschmidt echo starting from a Neel state \cite{Piroli2017, Piroli2018}. Our present work addresses this matter more generically by making a quantum many body Hamiltonian evolve from an initial state which involves all the quasiparticle eigenstates and eventually proves that the initial configuration can be manipulated in order to obtain a different type of dynamical quantum phase transition. The observation becomes more important because many experimental techniques that can be employed to realize the phenomena may not always cool down a system perfectly to its ground state.

In this paper, we consider an exactly solvable transverse Ising Hamiltonian which is paradigmatic of the class of integrable quantum many body system and also has been demonstrated to possess one kind of DQPT following a quench across QCP. We start from a generic state in the product form of superposition of the quasiparticle eigenstates. The unitary time evolution under a time-invariant Hamiltonian reveals that the DQPT occurs at infinite number of critical values of time, and these points of DQPT can be categorized into two types, which we call type A and type B. The latter type occurs only when the driving Hamiltonian resides in the ferromagnetic phase whereas the former takes place in both the phases. The appearance of one class of DQPT irrespective of the magnetic ordering of the Hamiltonian seems to provide an evidence of the fact that DQPT does not always imply the presence of a quantum critical point. The behavior of Fisher zeros are also different in these two cases. Type A singularities occur when a family of Fisher zeros are arranged in line along the imaginary axis and the corresponding DQPT occurs at the boundary points of each band \cite{Heyl-Review1}. On the other hand, a {\em line} of such zeros crosses the time axis in case of type B singularities and the DQPT occurs at the intersection point.

In the next section, we shall define the initial states and the expressions of the corresponding rate function in the form of integrals and in Sec.~\ref{behavior} we shall study the behavior of the rate function and obtain their singularities from the locations of Fisher zeros. The behavior of these quantities near the transition points will also be analyzed. In Sec.~\ref{occupancy} we find the occupancies of the eigenstates to explain the origin of the DQPTs. Finally we conclude with discussions in Sec.~\ref{conclusion}.

\section{Engineered initial state and the rate function}
\label{system}
\noindent
We consider a quantum Ising chain under constant transverse field. The $N$-spin Hamiltonian with periodic boundary condition is described by
\be   \mathcal{H}= -\sum_{i=1}^{N} s^{x}_{i} s^{x}_{i+1} - \Gamma\sum_{i=1}^{N} s^{z}_{i}  \label{H_def}  \ee
where $s^{x,z}$ are Pauli spin matrices. We have set the spin-spin interaction strength, $J=1$ and $\Gamma$ is the transverse field. The quantum critical point in this system is at $\Gamma_c = 1$ where the Hamiltonian exhibits an order-disorder transition.

The exact diagonalization of the Hamiltonian can be performed by mapping it onto that of non-interacting fermions with the help of Jordan-Wigner transformation. We can therefore write it as a Kronecker sum of commuting Hamiltonians ($\mathcal{H}_k$) of nonlocal free fermions of momenta $k$ \cite{LSM,Damski,BKC-book}
\bea \mathcal{H}_k &=& - 2 i \sin k \left[ a_k^{\dagger}a_{-k}^{\dagger} + a_k a_{-k} \right] \nonumber \\
&& - 2(\Gamma + \cos k) \left[ a_k^{\dagger}a_k + a_{-k}^{\dagger}a_{-k} - 1 \right] \label{Hk_def} \eea
where $k=(2n-1)\pi/N$ with $n=1,2,\cdots,N/2$. Each of these $\mathcal{H}_k$'s can be described by four basis states namely, $|00 \rangle_k$, $|11 \rangle_k$, $|10\rangle_k$ and $|01\rangle_k$, where the numbers in each basis signify the occupation status of the fermions having momenta $+k$ and $-k$ respectively. For example, the state $|11\rangle_k$ represents the state for which both the $+k$ and $-k$ modes are occupied by fermions, whereas for the state $|10\rangle_k$ the $-k$ mode is unoccupied. We neglect the parity-dependent boundary terms here as we perform all the calculations in the thermodynamic limit. The entire Hilbert space is therefore composed of the four-dimensional subspaces spanned by these basis states. For each $k$-mode, the ground state and the uppermost energy level correspond to the eigenstates having linear combinations of even-occupation states only. We write those eigenstates as
\begin{eqnarray}
|(\Gamma ,k)_{-}\rangle & = & i \cos \theta_k |11\rangle_k - \sin \theta_k |00\rangle_k \nonumber \\
|(\Gamma ,k)_{+}\rangle & = & i \sin \theta_k |11\rangle_k + \cos \theta_k |00\rangle_k 
\label{egstate}
\end{eqnarray}
with their eigenvalues $\mp\lambda_k = \mp 2\sqrt({\Gamma^2 +1 +2\Gamma \cos k})$ respectively, with $\tan \theta_k=-\sin k / \left[ \Gamma + \cos k + \lambda_k/2\; \right]$.

Two other eigenstates are the odd-occupation states, $|01\rangle_k$ and $|10\rangle_k$ and they have zero eigenvalues. In this fermionic representation, the ground state of the system can be written in terms of the quasiparticle ground states.\\

In our case we assume, primarily, that the system starts from such a configuration where each of the quasiparticle states is an arbitrary superposition of $|00\rangle_k$ and $|11\rangle_k$. For each momentum, such a state can, in general, be written as
\be |\psi_k(0)\rangle = \alpha_k(0) |11\rangle_k + \beta_k(0) |00\rangle_k \label{psi0even} \ee
with arbitrary combination of the coefficients satisfying the normalization condition. Starting from such an engineered configuration is somewhat similar to generating an initial configuration by using different quench protocols performed in most of the works on DQPT. As this state is spanned by the even-occupation basis states and the Hamiltonian does not couple it with the two other basis states $|01\rangle_k$ and $|10\rangle_k$, the later dynamics also remains confined in the space spanned by even occupation states only. If the system evolves under the time-invariant Hamiltonian $\mathcal{H}$ of Eq.~(\ref{H_def}), then $|\psi_k\rangle$ at any time $t$ may be obtained in the form
\begin{eqnarray}
|\psi_k(t)\rangle 
&=& \alpha_k(t) |11\rangle + \beta_k (t) |00\rangle \label{psiteven}
\end{eqnarray}
The coefficients can be calculated from the unitary evolution of $|\Psi_k\rangle$ under time-invariant $\mathcal{H}_k$.
\be  \left[ \begin {array}{l}
\alpha_k(t) \\
\beta_k(t)  \end{array}      \right]= 
\left[ \begin {array}{rl}
\mathcal{U} & \mathcal{V} \\
-\mathcal{U} & \mathcal{U}^\ast  \end{array}      \right]
 \left[ \begin {array}{l}
 \alpha_k \\
 \beta_k
 \end{array}  \right] 
\ee
where $\mathcal{U}=\cos^2\theta_k \exp(i\lambda_k t) + \sin^2\theta_k \exp(-i\lambda_k t)$ and $\mathcal{V}=\sin2\theta_k\sin\lambda_k t$
with $t$ is scaled by $\hbar$ and the zeros in the initial coefficients have been dropped for the sake of convenience.\\

The initial state of the whole system is represented as
\be |\Psi (0) \rangle = |\psi_{k_1} (0) \rangle |\psi_{k_2} (0) \rangle \cdots | \psi_{k_{N/2}} (0) \rangle \label{Psi_product} \ee
where $k_n = (2n-1)\pi/N$ are the wave-vectors, as defined below Eq. (\ref{Hk_def}).  
This state can be written in term of creation and annihilation operators applied to a vacuum state $|{\mathcal V}\rangle$, which contains no fermion at any site,  as
\be |\Psi\rangle = {\mathcal Q}_{N/2} {\mathcal Q}_{N/2-1} \cdots {\mathcal Q}_2 {\mathcal Q}_1 |{\mathcal V}\rangle \label{O_product} \ee
where
\be {\mathcal Q}_ n = \alpha_{k_n} \, a_{k_n}^{\dagger} a_{-k_{n}}^{\dagger} + \beta_{k_n}  \, {\mathbf{1}}   \label{O_def} \ee
Note that the sites in the lattice are arranged as $k_1, -k_1, k_2, -k_2, \cdots , k_{N/2}, -k_{N/2}$ from left to right. 
The wave function at time $t$, namely $|\Psi(t)\rangle$ will also be of the form of Eq. (\ref{Psi_product}). This enables us to express the Loschmidt amplitude and the return probability into the product form in momentum space  
\be
\mathcal{G}(t) = \prod\limits_{k} g (k,t) \label{G-even}
\ee
and
\be
\mathcal{L}(t) = \prod\limits_{k} \varrho (k,t) \label{L-even}
\ee
where $g(k,t) = \langle \psi_k (0) | \psi_k (t)\rangle$ and $\varrho (k,t) = |\langle \psi_k (0) | \psi_k (t)\rangle|^2$. Consequently the rate function determined in the thermodynamic limit takes the form
\be \mathcal{R} (t) 
= - \dfrac{1}{\pi}\int\limits_0^{\pi} \log \varrho (k,t)\; dk \label{r-even} \ee
where $\varrho(k,t)=\cos^2 \lambda_k t + (|\alpha_k|^2 - |\beta_k|^2)^2 \cos^2 2\theta_k \sin^2 \lambda_k t$.

The non-analyticities in $\mathcal{R}(t)$ could be found from $\varrho = 0$ only, but we shall explain later that $\varrho = 0$ does not always imply the non-analyticity in $\mathcal{R}(t)$, which in another sense is the reason why we consider only the two end-points of the line of Fisher zeros along time axis as critical times of DQPT \cite{Heyl-Review1}.\\

\begin{figure*}
\subfloat[]{
  \includegraphics[scale=0.33]{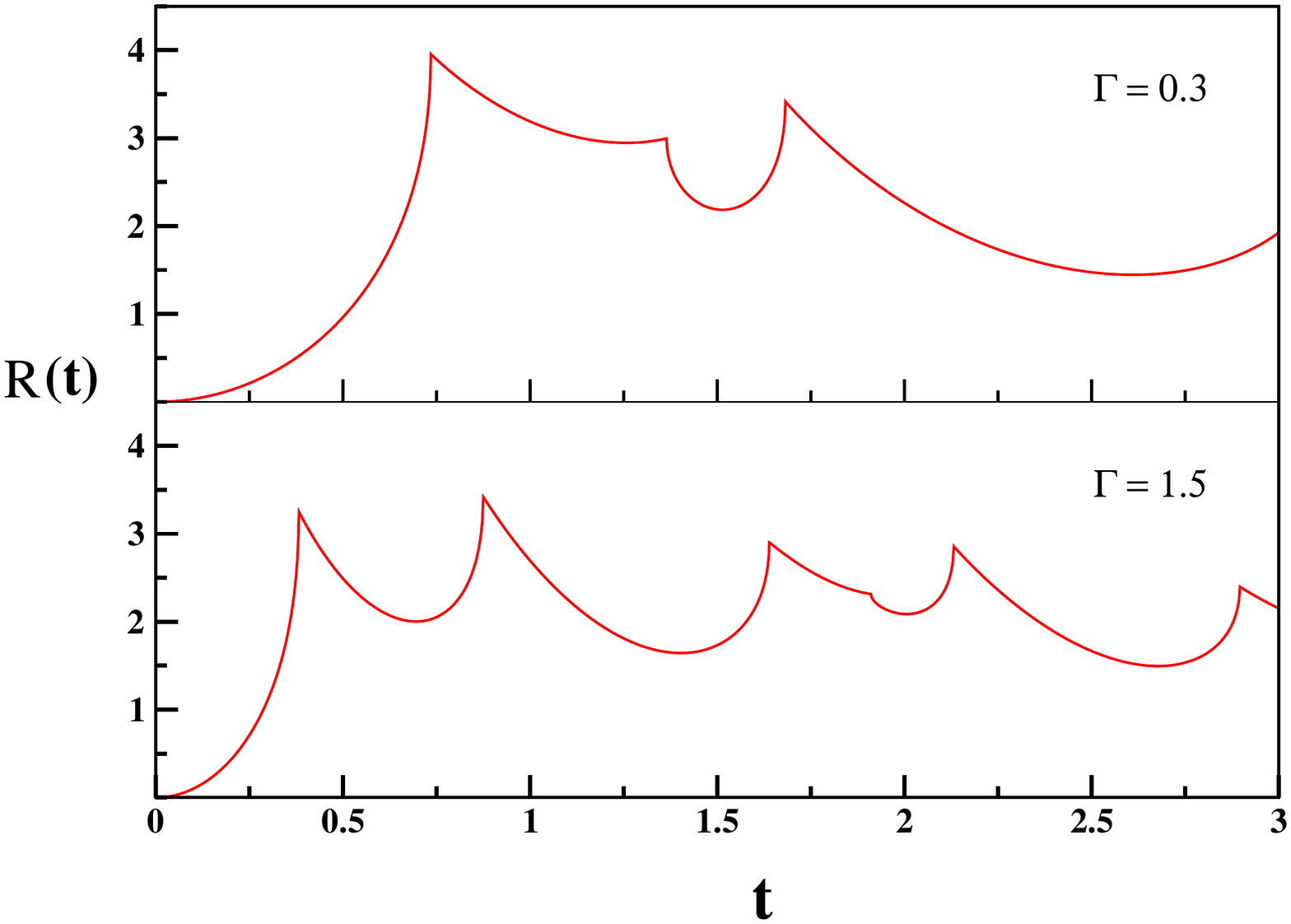}
  \label{typeA}
}
\hspace*{-10mm}
\subfloat[]{
  \includegraphics[scale=0.33]{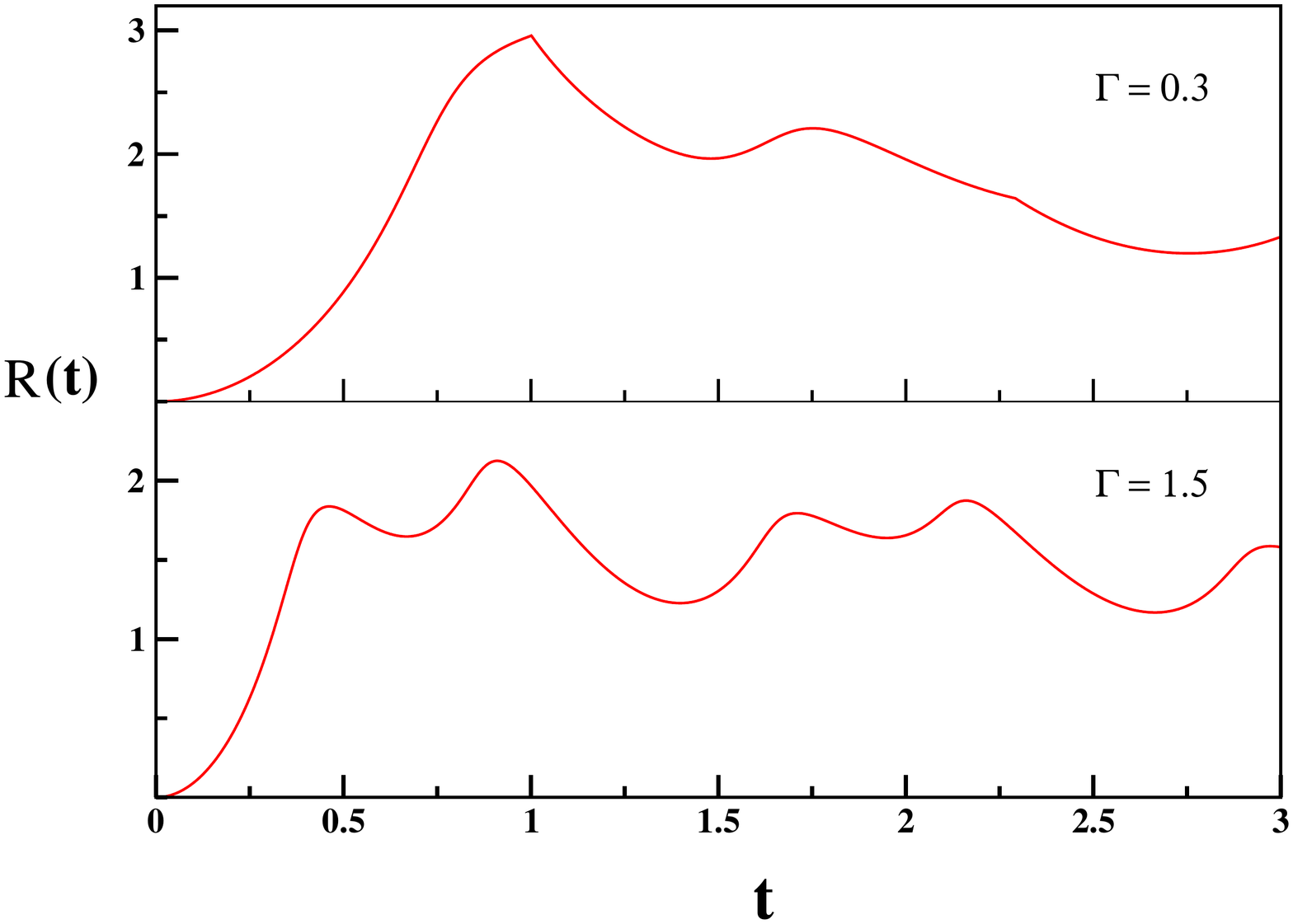}
  \label{typeB}
}

\caption{Non-analyticities shown by the rate function when $\mathcal{E}>\mathcal{O}$: (a) Type A kinks are shown when $|\alpha_k| = |\beta_k|$ in both the cases of $\Gamma<1$ and $\Gamma>1$; (b) Kinks of type B appear when $|\alpha_k|\neq|\beta_k|$ and $\Gamma<1$. These kinks do not appear for $\Gamma>1$. We have taken $|\alpha_k|^2 = 0.5$ and $|\beta_k|^2 = 0.25$ in this case.}
\label{rgen-fig}
\end{figure*}

The rate function derived here shows two types of non-analyticities as a function of time. We identify them as follows:\\
{\bf Type A}: For $|\alpha_k|=|\beta_k|$, we find repeated kinks along time-axis, which appear for all values of $\Gamma$ and hence has no connection with the QCP of the Hamiltonian (Fig.~\ref{typeA}).\\
{\bf Type B}: For $|\alpha_k|\neq |\beta_k|$, another type of non-analyticity exists but this exists only when the Hamiltonian is in ordered (along $x$-direction) phase, i.e, $\Gamma<1$ (Fig.~\ref{typeB}).\\

Now we construct a more general form of initial state which involves all the quasiparticle eigenstates for each momentum.
\be |\psi_k(0)\rangle = \alpha_k |11\rangle_k + \beta_k |00\rangle_k + \gamma_k |10\rangle_k + \delta_k |01\rangle_k \label{psi0}  \ee
The state of the whole system is still of the form of Eqs. (\ref{Psi_product}) and (\ref{O_product}) with the operator ${\mathcal Q}_ n$ given by 
\be {\mathcal Q}_ n = \alpha_{k_n} \, a_{k_n}^{\dagger} a_{-k_{n}}^{\dagger} + \beta_{k_n}\;{\mathbf{1}} + \gamma_{k_n}  \sigma_n \, a_{k_n}^{\dagger} 
+ \delta_{k_n}  \sigma_n \, a_{-k_n}^{\dagger} \ee
Here, $\sigma_1 = 1$ and for $n > 1$,
\[ \sigma_n = \prod_{j=1}^{n-1} \left(1-2a_{k_j}^{\dagger} a_{k_j}\right)  \left(1-2a_{-k_j}^{\dagger} a_{-k_j}\right) \]
\noindent
The coefficients $\gamma_k$ and $\delta_k$ of Eq.~(\ref{psi0}) will not change with time because of the corresponding eigenstates having zero-eigenvalues. One can note that Eq.~(\ref{G-even}) and (\ref{L-even}) are also valid for the general case of Eq.~(\ref{psi0}). Also, the Hamiltonian conserves the total probabilities of even and odd occupation for each $k$-mode separately, which enables us define $\mathcal{E} = |\alpha_k(t)|^2 + |\beta_k(t)|^2 = |\alpha_k|^2 + |\beta_k|^2$ and $\mathcal{O} = |\gamma_k(t)|^2 + |\delta_k(t)|^2 = |\gamma_k|^2 + |\delta_k|^2$ with $\mathcal{E} + \mathcal{O} = 1$. Note that $\mathcal{E}$($\mathcal{O}$) is the sum for the even(odd) parity states of each mode and is not the same as the total parity of the eigenstates of $\mathcal{H}$.

The return probability and the rate function can be written as those in Eqs.~(\ref{L-even}) and (\ref{r-even}) with
\be
\varrho(k,t) = (\mathcal{O} + \mathcal{E}\cos \lambda_k t)^2 + (|\alpha_k|^2 - |\beta_k|^2)^2 \cos^2 2\theta_k \sin^2 \lambda_k t
\label{rho} \ee 
 
We now discuss the behavior of the rate function and the Fisher zeros which are the keys to investigate the DQPT of the system.

\section{Behavior of the rate function and Fisher zeros}
\label{behavior}

\noindent
The necessary condition for non-analyticity in $\mathcal{R}(t)$ is $\varrho (k,t)=0$. It is evident from Eq.~(\ref{rho}) that, the rate function shows no singularity for $\mathcal{E}<\mathcal{O}$ as the first term always remains non-zero. On the other hand, for $\mathcal{E}\geq\mathcal{O}$, two different types of non-analyticities are shown by $\mathcal{R}(t)$.

If we consider first the case of $|\alpha_k|=|\beta_k|$, the generalized expression for $\varrho (k,t)$ reduces to $\varrho(k,t) = (\mathcal{O} + \mathcal{E}\cos \lambda_k t)^2$ and the rate function becomes
\be \mathcal{R}(t) = - \dfrac{1}{\pi}\int\limits_0^{\pi}  \log (\mathcal{O} + \mathcal{E}\cos \lambda_k t)^2 \; dk \label{rA} \ee
From the integrand of Eq.~(\ref{rA}) it may seem at first sight that if we are in the regime $\mathcal{E}\geq\mathcal{O}$, then for any $t$, there will be some value of $k$ for which $\cos\lambda_k t = -\mathcal{O}/\mathcal{E}$ and the integral will show a non-analyticity. But in practice, $\mathcal{R} (t)$ shows kinks at some particular points ($t_c$) on the time axis because the integral of such type is dominated by contributions from the band edge \cite{Andraschko}. In other words, if we define those $k$-values as critical modes ($k_c$) which are responsible for the non-analyticity on $\mathcal{R}(t)$ then, for this type of DQPT, $k_c = 0$ and $\pi$.

The rate function has another type of non-analyticity for $|\alpha_k|\neq|\beta_k|$,  when we have $\cos 2\theta_k=0$ and $\cos\lambda_k t=-\mathcal{O}/\mathcal{E}$ simultaneously in Eq.~(\ref{rho}). In this case the critical momentum mode which make $\cos 2\theta_k$ vanish is given by 
\be k_c = \cos^{-1}(-\Gamma) \label{kb}\ee
Noting that $\lambda_{k=k_c}=2\sqrt{(1-\Gamma^2)}$, the non-analyticity occurs at 
\be t_c = \dfrac{\cos^{-1}(-\mathcal{O}/\mathcal{E})}{2\sqrt{1-\Gamma^2}} \label{tb} \ee
The equations (\ref{kb}) and (\ref{tb}) indicate that this type of non-analyticity will not exist for $\Gamma>1$ i.e., the quantum critical point.

\begin{figure}
\hspace*{-5mm}
\includegraphics[scale=0.35]{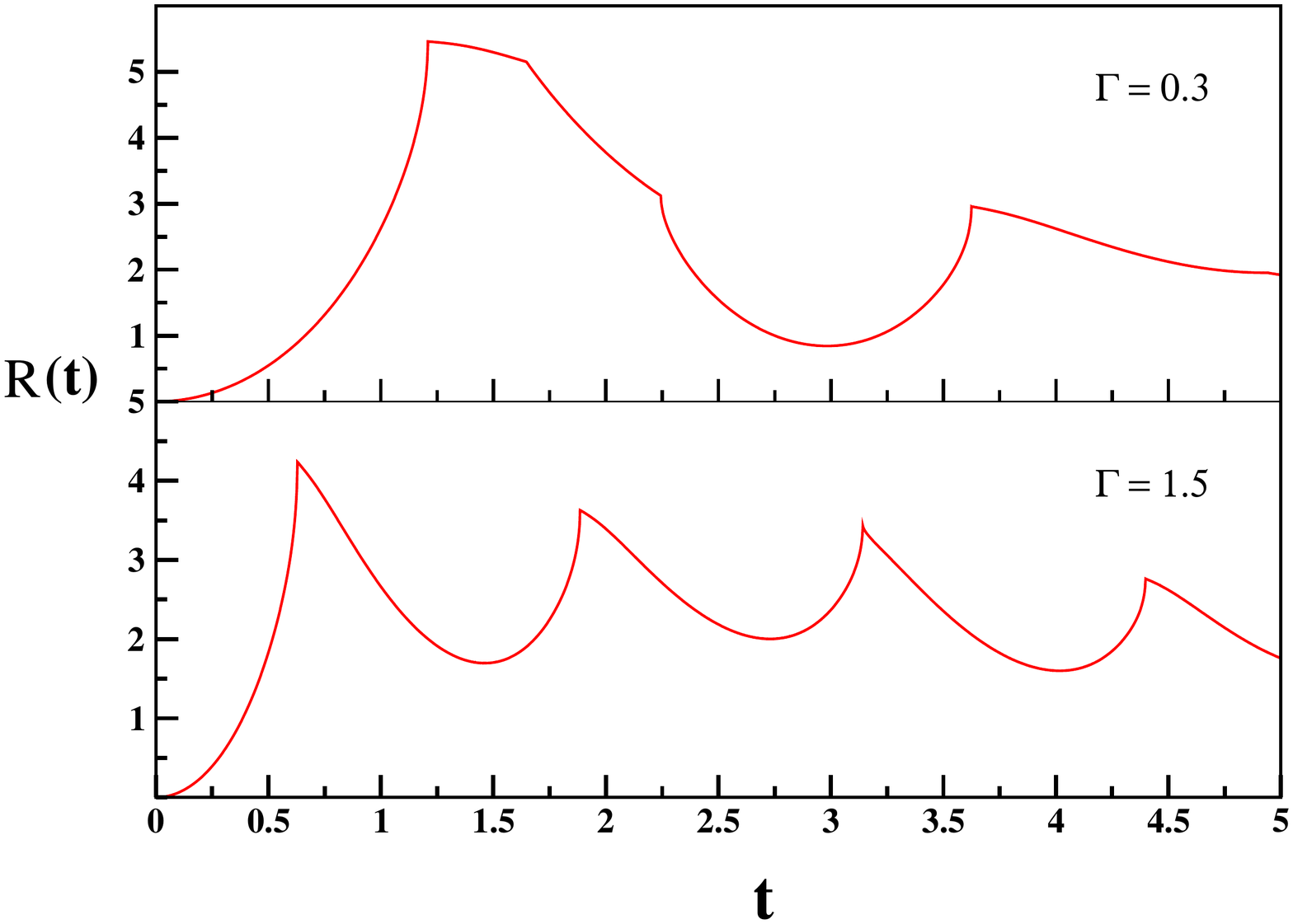}
\caption{Non-analyticities shown by the rate function when $\mathcal{E}=\mathcal{O}$. The upper panel shows both type A and type B non-analyticities for $\Gamma<1$. Starting from $t=0$, the $1$st, $3$rd and $4$th kinks are of type A and the $2$nd kink is of type B. The lower panel shows non-analyticities of only type A for $\Gamma>1$. In both the cases $|\alpha_k|^2 = 0.35$}
\label{typeAB}
\end{figure}

However for a special case when $\mathcal{E}=\mathcal{O}=1/2$, the rate function possesses type A kinks even for unequal coefficients. 
\bea \mathcal{R}(t) &=& - \dfrac{1}{\pi}\int\limits_0^{\pi} dk \, \log\left[ \cos^2 \frac{\lambda_k}{2} t \left\{\cos^2 \frac{\lambda_k}{2} t \right. \right. \nonumber \\
&& \left.\left.+ 4(|\alpha_k|^2 - |\beta_k|^2 )^2 \cos^2 2\theta_k \sin^2 \frac{\lambda_k}{2} t \right\} \right] \eea
Fig.~\ref{typeAB} shows the kinks occurred in the rate function for this case.\\

The critical exponent for the non-analyticity of rate function may be defined as
\be {\mathcal R}(t) = {\mathcal A} +  {\mathcal B} (t - t_c)^{\nu} \ee
for small values of $(t-t_c)$, where $ {\mathcal A}$ and $ {\mathcal B}$ are independent of time. Generally the DQPT has the critical exponent $\nu=1$ \cite{Heyl-Review1}. However different universality classes of DQPT have been reported recently with different critical exponents \cite{Halimeh,Wu}. In our case of type B non-analyticity, the critical exponent can be easily derived to be $\nu = 1$. However, for type A, the exponent is atypically different in two sides.

The derivative of the rate function diverges when $t \to t_c^{-}$ . From Eq.~(\ref{rA}) we obtain

\be \partial_t \mathcal{R} = \dfrac{2}{\pi}\int\limits_0^\pi \dfrac{\mathcal{E}\lambda_k \sin\lambda_k t}{g(k,t)}\;dk \label{rA-deriv}\ee
where $g(k,t) = \mathcal{O} + \mathcal{E}\cos\lambda_k t$. It is easily seen that a zero value of $g(k,t)$ at some $t$ does not always imply divergence of the integral at the left. Thus, suppose for some $k=k_c$, one finds some $t=t_c$ for which $g(k,t)=0$. Since $\lambda_k$ decreases monotonically with $k$, at $k$ slightly above (below) $k_c$, the function $g(k,t)$ has a small positive(negative) value and the integrand has a large positive(negative) value. These large values cancel out, leading a finite value of the integral. The situation is however different for $k_c = 0 \text{  or } \pi$.

For $k_c=0$, one has $\mathcal{O} + \mathcal{E}\cos [ 2(\Gamma + 1)t_c ] = 0$, since $\lambda_{k=0}=2(\Gamma+1)$. For $t=t_c - \epsilon$, the function $g(k,t)$ does not vanish for any $k$ but decreases monotonically to a small value at $k=0$, remaining positive always. Hence the integral diverges. For $t=t_c + \epsilon$, the function $g(k,t)$ vanishes at some small positive value of $k$, say $k_0$, and attains small negative(positive) values for $k$ slightly less(greater) than $k_0$. Hence there is a cancellation of large positive and negative values of the integrand and the integral does not diverge. It can be shown similarly that for $k_c=\pi$, the integral remains finite just below $t_c$ but diverges just above $t_c$. 

Let us now calculate the exponent of divergence of the derivative near $t_c$. Noting that if we consider the kinks of Eq.~(\ref{rA}) arising from $k_c = 0$, the derivative $\partial_t {\mathcal R}$ of Eq.~(\ref{rA-deriv}) diverges at $t=t_c^{-}$ and remains finite at $t=t_c^{+}$, indicating a non-analyticity at $t=t_c$ and the dominant contribution to the integral in Eq.~(\ref{rA-deriv}) comes from small values of $k$ (say, from the region $0 < k < \delta$). Let us assume first $t = t_c -\epsilon$, where $\epsilon$ a positive but small quantity. We may then expand $\lambda_k$ about $k=0$ and retain the leading order terms to obtain
\begin{eqnarray}
\partial_t \mathcal{R} &=& \dfrac{4\mathcal{E}(\Gamma + 1)}{\pi}  \int\limits_0^{\delta} \dfrac{\sin \varphi }{\mathcal{O} + \mathcal{E}\cos \varphi} \; dk
\end{eqnarray}
where $\varphi$ is a function of $k$: $$\varphi = 2(\Gamma + 1)t_c - 2(\Gamma + 1)\epsilon - \frac{\Gamma t_c}{(\Gamma + 1)}k^2$$. In the small $\epsilon$ limit, the above integral gives
\be \partial_t \mathcal{R} = \dfrac{(\Gamma + 1)^{3/2}}{\sqrt{\Gamma\cos^{-1}(-\mathcal{O}/\mathcal{E})}} \cdot \dfrac{1}{\sqrt{t_c - t}} + \dfrac{4(\Gamma + 1)\mathcal{O}\delta}{\pi\sqrt{\mathcal{E}^2 - \mathcal{O}^2}} \ee
where we have put $2(\Gamma + 1)t_c = \cos^{-1}(-\mathcal{O}/\mathcal{E})$. This proves that $\partial_t \mathcal{R}$ diverges as $t \to t_c^-$. A similar situation occurs for $k_c = \pi$ as well. Such behavior of the derivative makes the rate function approach the critical point as $(t-t_c)^{1/2}$, making the critical exponent, $\nu=1/2$ in this case. On the other hand, $\nu = 1$ for $t \to t_c^+$ as $\partial_t \mathcal{R}$ remains finite in this regime.\\

\subsection*{Fisher zeros}
\noindent
The appearance of the above DQPTs can be extracted from the Fisher zeros of the dynamical partition function also. The dynamical partition function for the system starting from the non-eigenstate (\ref{psi0}) and evolving under a constant Hamiltonian can be written as
\bea \mathcal{Z}(z) &=& \prod\limits_{k} \left[ \mathcal{O} + \left( |\alpha_k|^2 \cos^2 \theta_k + |\beta|^2 \sin^2 \theta_k \right)e^{\lambda_k z} \right. \nonumber \\
&& \left. \;\;\;\;\;+ \left( |\alpha_k|^2 \sin^2 \theta_k + |\beta|^2 \cos^2 \theta_k \right)e^{-\lambda_k z} \right] \label{gz} \eea
These Fisher zeros of $\mathcal{Z}(z)$ produces closely spaced points for a finite-size systems that would form continuous lines or surfaces in the thermodynamic limit and when they cross the imaginary (real time) axis, we expect to observe DQPTs in the real-time behavior of the system. From Eq.~(\ref{gz}) we obtain a family of lines of Fisher zeros labeled by some integer $n$:
\be z_n(k) = \dfrac{1}{\lambda_k}\left[\; \ln r + i(2n + 1)\pi \; \right] \label{zn} \ee
where $r = \dfrac{\mathcal{O} \pm \sqrt{\mathcal{O} - \mathcal{E} + (|\alpha|^2 - |\beta|^2)^2 \cos^2 2\theta}}{\mathcal{E} + (|\alpha|^2 - |\beta|^2) \cos 2\theta}$\\

\begin{figure*}
\subfloat[]{
  \includegraphics[scale=0.48]{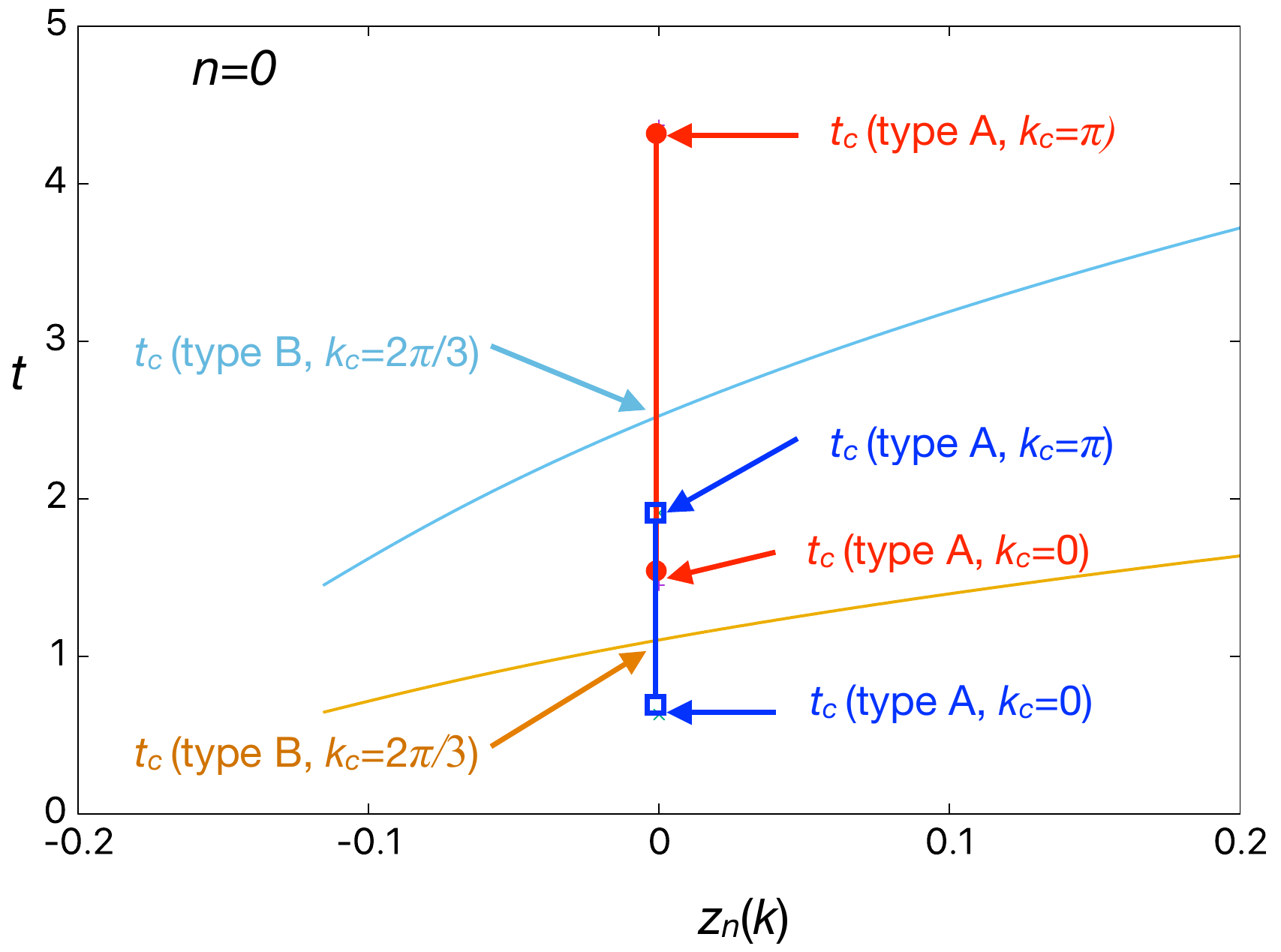}
  \label{fz1}
}
\subfloat[]{
  \includegraphics[scale=0.48]{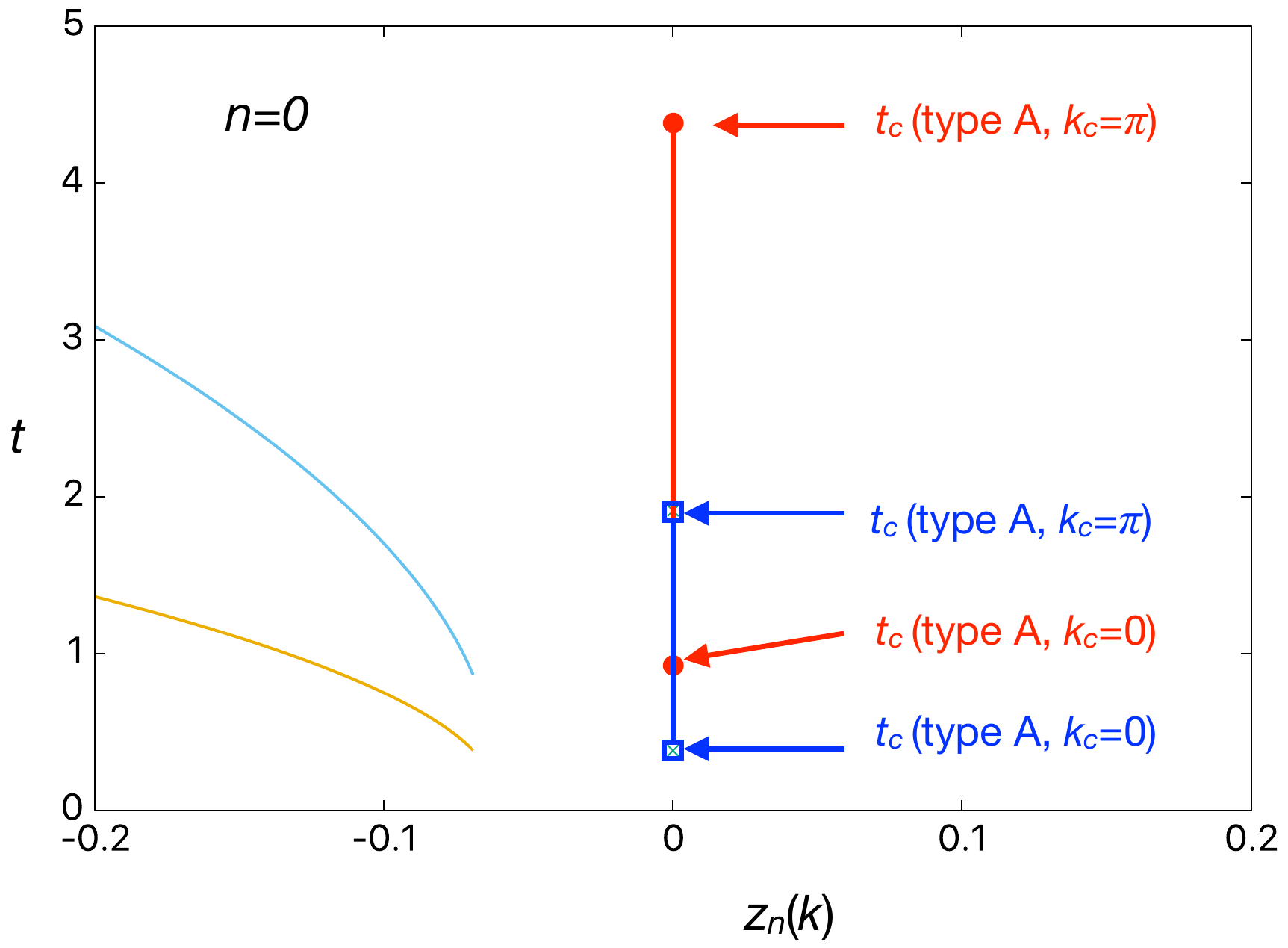}
  \label{fz2}
}
\caption{Fisher zeros in complex plane for different cases when $\mathcal{E}>\mathcal{O}$. (a) Both types of DQPT appear when $\Gamma = 0.5$. For $|\alpha_k|=|\beta_k|$ the family of Fisher zeros lie along the time axis. The DQPT appears at the band edges for each $n$. For $|\alpha_k|\neq|\beta_k|$ the lines of Fisher zeros cut the time axis at different points where the DQPTs occur; (b) Only type A DQPT occurs when $\Gamma = 1.5$. In this case, for $|\alpha_k|\neq|\beta_k|$ (here $|\alpha_k|^2 = 0.5$ and $|\beta_k|^2 = 0.25$) the lines of Fisher zeros do not cut the time axis and type B DQPT cannot be obtained. Moving along the time axis from $t=0$, the first three points of type A DQPT correspond to the $1$st, $2$nd and $4$th kinks shown in the lower panel of Fig.~\ref{typeA}. The other kinks of that curve in Fig.~\ref{typeA} actually correspond to Fisher zeros for different $n$.}
\label{fz-fig}
\end{figure*}

The behavior of Fisher zeros on the complex plane give a good indication to the characteristics of the DQPT. It can be shown from (\ref{zn}) and Fig.~\ref{fz-fig} that $z_n$ cannot touch the imaginary axis when $\mathcal{E}<\mathcal{O}$ which implies the fact that DQPT cannot occur in this case. On the other hand, for $\mathcal{E}>\mathcal{O}$, two cases arise: (A) the Fisher zeros fall on the imaginary axis for $|\alpha_k|=|\beta_k|$ and (B) for $|\alpha_k|\neq |\beta_k|$, the lines of zeros cut or do not cut the time axis depending on whether the transverse field is below or above the quantum critical point. 

In the former case, the DQPT can occur only at the boundary points of the line of Fisher zeros on the imaginary axis. The analysis of $\mathcal{R}(t)$ and $\partial_t\mathcal{R}$ has already established that those points correspond to the band-edges in the quasiparticle picture.

\section{Occupancy of eigenstates}
\label{occupancy}

\noindent
The origin of the two types of DQPT can be explained in the light of the occupancy of the eigenstates of the Hamiltonian. The time evolution of the system is entirely governed by two eigenstates, $|(\Gamma, k)_{-}\rangle$ and $|(\Gamma, k)_{+}\rangle$ for each $k$-mode (Eq.~\ref{egstate}). We define the occupancy of these states in the initial state as

\bea
n_k^g = |\langle (\Gamma, k)_{-}|\psi_k (0)\rangle |^2 = |\alpha_k|^2 \cos^2 \theta_k + |\beta_k|^2 \sin^2 \theta_k \nonumber \\
n_k^e = |\langle (\Gamma, k)_{+}|\psi_k (0)\rangle |^2 = |\alpha_k|^2 \sin^2 \theta_k + |\beta_k|^2 \cos^2 \theta_k
\eea
\noindent
Each $k$-component of the Loschmidt amplitude defined in Eq.~(\ref{G-even}) is given by
\be
g(k,t) = \mathcal{O} + (n_k^g + n_k^e)\cos\lambda_k t + i (n_k^g - n_k^e) \sin \lambda_k t
\ee
\noindent
For both types of DQPT, ${\rm{Im}}\left[ g(k,t) \right]$ vanishes when these two occupancies become equal. The critical modes ($k_c= 0$ and $\pi$ for type A and $k_c=\cos^{-1}(-\Gamma)$ for type B) make the real part of $g(k,t)$ vanish at the corresponding critical times.

The equal occupancy is interpreted as the maximum mixing of the quasiparticle eigenstates which can again be thought of as forming an infinite temperature state \cite{Heyl-Review1}. The time evolution of a generic state ensures the fact that irrespective of performing a sudden quench or crossing a QCP in such systems, the equal occupancy of the eigenstates in $k$-space that control the dynamics becomes the crucial necessary condition for DQPT.

\section{Conclusion and outlook}
\label{conclusion}

We have studied the time evolution of a generic state of a transverse Ising Hamiltonian. This simple unitary evolution identifies two distinct classes of DQPT, both of which depend crucially on the initial state of the system. The first type of non-analyticity has not been detected in any quench protocol studied in this class of system so far and it has no connection with the quantum critical point of the Hamiltonian. The existence of the second type depends on the magnetic phase of the Hamiltonian. Some particular critical momentum modes are responsible for the non-analyticities to occur at different critical times. The critical exponent is found to be $1$ for type B non-analyticity, but it is surprisingly non-identical in two sides of type A due to the divergence of slope only on one side of the critical time. Such behavior of the critical exponent is contradictory to the so called universality of the DQPT. Moreover, the attributes observed here indicate the fact that DQPT does not always follow a quantum phase transition as well. The equal occupancy of the quasiparticle eigenstates is rather found to be the necessary condition for this phenomena.

We can therefore classify DQPT in quantum many body system as two types: one which is independent of the occurrence of QPT and the other which is associated to it. However, there are more questions to be answered. First of all, more investigation on different integrable and non-integrable systems are needed especially to explore the possibility of defining the new type of DQPT in a generic quantum many body system when the quasi-particle representation is absent. An important question is how a QPT influences a DQPT at all and whether the underlying Kibble-Zurek mechanism has any role in it. Another question which eventually arises is whether the DQPT can be designated as a phase transition or not, especially  when the characterization is hindered by its sensitivity towards the initial configuration and no universal behavior is found to exist. More work on these issues is in progress.  \\

\section*{Acknowledgements}
\noindent
This research was encouraged by the International Centre for Theoretical Sciences (ICTS) during a visit of SD for participating in the program - Indian Statistical Physics Community Meeting (Code: ICTS/ispcm2019/02).

\end{document}